%Paper: hep-th/9306087
%From: Lev Vaidman <LEV@taunivm.tau.ac.il>
%Date: Fri, 18 Jun 93 11:41:12 IST

 %macropackage=phyzzx

\PHYSREV

\titlepage \nopubblock \title{Causality constraints on nonlocal quantum
measurements} \author{Sandu Popescu}

\address{ Service de Physique Th\'eorique \break  Universit\'e Libre de
Bruxelles \break Campus Plaine CP 225, Bld du Triomphe 1050 Bruxelles,
BELGIUM}

\andauthor{Lev Vaidman}

\address{ School of Physics and Astronomy \break Raymond and Beverly
Sackler
Faculty of Exact Sciences \break Tel-Aviv University, Tel-Aviv, 69978
ISRAEL \break BITNET: LEV@VM.TAU.AC.IL}

\abstract

Consequences of relativistic causality for measurements of nonlocal
characteristics of composite quantum systems are investigated.  It is
proved that verification measurements of entangled states necessarily
erase local information.  A complete analysis of measurability of
nondegenerate spin operators of a system of two spin-1/2 particles is
presented.  It is shown that measurability of certain projection
operators which play an important role in axiomatic quantum theory
contradicts the causality principle.
\vfill
hep-th\9306087 ,  TAUP 2011-92.

\endpage

\def \ra {\rangle}  \def \pop {$|\psi _{\perp}\rangle$}
\def
\po {$|\psi _0\rangle$} \def \p {$|\psi \rangle$} \def \pp {$|\psi'
\rangle$}
\def \ppp {$|\psi'' \rangle$} \def \P {|\psi \rangle} \def \PP {|\psi'
\rangle}
\def \PPP {|\psi'' \rangle} \def \PO {|\psi _0\rangle} \def \u {\vert
\uparrow}
\def \d {\vert \downarrow} \def \n {{1\over \sqrt 2}}

{\bf \chapter {Introduction }}

As early as 1931, Landau and Peierls \Ref\LP{L.~Landau and R.~Peierls
\journal Z. Physik & 69 (31) 56.} showed that relativistic causality
imposes new restrictions on the process of quantum measurement.
Although some of their arguments were not precise, it was commonly
accepted that we cannot instantaneously measure nonlocal properties
without breaking relativistic causality.  It was only in 1980 that
Aharonov and Albert \Ref\AA{ Y.~Aharonov and D.~Albert \journal Phys.
Rev.& D21 (80) 3316, and {\bf D24} (1981), 359.} showed that there are
nonlocal variables which
can be instantaneously measured without contradicting relativistic
causality.  The work of Aharonov and collaborators
\Ref\AAV{Y.~Aharonov, D.~Albert, and L.~Vaidman \journal Phys.  Rev.&
D34, (86) 1805.} was mainly devoted to finding explicit methods for
performing instantaneous nonlocal measurements.  Here we derive
general properties of all such measurements.  Our main result is that
instantaneous nonlocal measurement invariably disturbs the measured
system in such a way that all local information (except for that which
is related to degrees of freedom not involved in the measurement) is
erased.  That is, when nonlocal measurements are performed on an
ensemble of systems, there are limits on the amount of information
about the initial state contained in the final state of the ensemble.

Section 2 is devoted to defining the framework of our research.  In
Section 3 we discuss the causality principle and we derive useful
equalities which follow solely from causality.  In Section 4 we
define state verification measurement, and, based on the requirement
of reliability of the measurement, we prove two theorems about
necessity of erasing local information in verification measurements of
entangled states.  Section 5 is devoted to application of the derived
results to analysis of the measurability of spin operators of a system
of two spin-1/2 particles.  It includes a complete analysis of the
measurability of nondegenerate operators of this system.  In Section 6
we investigate consequences of our results on axiomatic quantum
theory showing that certain {\it ideal measurements of the first
kind}\Ref\Pir{C.~Piron, {\it Foundations of Quantum Physics}, W.A.
Benjamin Inc. (1976) p.68.} cannot be performed without contradicting
the causality principle.  A summary of our results concludes the paper
in Section 7.

{\bf \chapter {The General Framework}}

Our present study is restricted to the framework which is generally
used for investigating causality constraints on quantum measurements.
It was first applied by Bohm and Aharonov \Ref\BA{D.~Bohm and
Y.~Aharonov \journal Phys.Rev. &108 (57) 1070.} and later by Bell
\Ref\Bell{J.~S.~Bell \journal Physics &1 (64) 195.} for analyzing the
Einstein-Podolsky-Rosen argument; and this was the framework in which
the first measurable nonlocal variables were found\rlap.\refmark{\AA}\
That
is, we consider quantum systems which consist of two distinguishable
parts, each localized in a different region of space.  We take each
region small enough to allow neglecting causality restrictions inside
it, but much bigger than a Compton wave length, so we can neglect
relativistic effects such as pair creation.  The causality principle
enters at the scale of distances between the widely-separated parts of
the system, while locally we can use the formalism of nonrelativistic
quantum mechanics.  For example, we shall consider a system of two
spin-1/2 particles located in remote space regions.  Spin components
of each particle can be measured by using a Stern-Gerlach apparatus,
and these measurements are described by nonrelativistic quantum
mechanics.

In this work we study {\it instantaneous measurements of nonlocal
properties}.  Following von Neumann, any measurement can be considered
as having three stages.  The first stage is a preparation of the
measuring device.  The second stage is an interaction between the
measured system and the measuring device.  As the result of this
interaction the final state of the measuring device will contain
information about the initial state of the system.  In the third stage
this information is read by observers.

By an {\it instantaneous} measurement we do not mean that some
observer can instantaneously find out the result.  Only the second
stage of the measurement, the interaction between the measuring device
and the system, must be instantaneous (\ie very short).  The
measurement as a whole may take a much longer time: the measuring
device may have had to be prepared a long time before the interaction
and it may take a long time to recover the result.

The measurements considered here are designed to determine nonlocal
properties of systems.  Nevertheless, the interaction between the
measuring device and the measured system need not be nonlocal, since
it is possible\refmark{\AA} to measure {\it nonlocal} properties via
{\it local} interactions.  However, in the present work we do not make
any particular assumptions about the interaction between the system
and the measuring device, apart from unitarity of the time evolution.

{\bf \chapter {The Causality Principle}}

The causality principle states that observers situated near widely
separated parts of a system cannot communicate with one another with
superluminal velocity: local interactions performed in one part of the
system could not affect the probabilities of the outcomes of local
measurements performed on the other part of the system outside the
light cone.  Consider now the following situation: Our system,
consisting of two separate parts ``1" and ``2", is prepared,
initially,
in a state \p . At time $t_0-\epsilon$ (with $\epsilon c$ small
compared
to the distance between the two parts of the system) some local
interaction is performed, say, on part 2 of the system.  This
interaction is described by a unitary transformation $U^{(2)}$.  At
time $t_0$ a nonlocal measurement is performed.  More exactly, at
$t_0$ the measuring device interacts with the system and this leads to
a unitary transformation {\twelveit U} of the state of the
composite: the system and the measuring device.  At time
$t_0+\epsilon$ a measurement of a local observable $A^{(1)}$ is
carried out on part 1. From the causality principle it follows that
the probability for any particular outcome of any local measurement,
say $A^{(1)}=a$, is independent of the local action on part 2,
$U^{(2)}$.

Let us denote the probability for the result $A^{(1)}=a$ of the local
measurement in part 1 at time $t_0+\epsilon$, provided the state of
the system
immediately before the nonlocal measurement performed at $t_0$ was \p, by
$p(\psi)$:
$$
p(\psi) \equiv prob \Bigl( A=a_{{\rm~at~}t_0+\epsilon} \Bigm|~|\psi \ra
_{{\rm ~at~}t_0-\epsilon}; {\rm ~nonlocal \ measurement}_{{\rm ~at~} t_0}
\Bigr).\eqno(1)
$$
In this compact notation the causality principle yields:
$$
 p(U^{(2)}\psi) = p(\psi).\eqno(2)
$$

The probability for a given outcome of a measurement is equal to the
expectation value of the projection operator onto the corresponding
subspace.
Let ${\bf P}_a^{(1)}$ denote the projection operator onto the subspace
corresponding to $A^{(1)}=a$. Then,
$$
 p(\psi) = \langle \phi | \langle \psi | U^{\dagger} {\bf P}_a^{(1)} U |
\psi
\ra |\phi \ra, \eqno(3)
 $$
 where $|\phi \ra$ is the initial state of the measuring device,
and
where the unitary transformation {\it U} describing the nonlocal
measurement
acts on both states, that of the system and that of the measuring device.
In
this more explicit notation the causality principle (2) becomes
$$
 \langle \phi | \langle \psi | {U^{(2)}}^\dagger U^\dagger {\bf
P}_a^{(1)} U
U^{(2)} | \psi \ra |\phi \ra =  \langle \phi | \langle \psi | U^\dagger
{\bf
P}_a^{(1)} U | \psi \ra |\phi \ra .\eqno(4)
 $$
By applying Eq.(4) to the states $\vert \psi_1\ra+\vert \psi_2\ra$ and
$\vert\psi_1\ra+i\vert\psi_2\ra$, where $\vert\psi_1\ra$ and
$\vert\psi_2\ra$ are arbitrary, we easily obtain the
following generalization of (4):
$$
 \langle \phi | \langle \psi \mathstrut_2| {U^{(2)}}^\dagger U^\dagger
{\bf
P}_a^{(1)} U U^{(2)} | \psi _1\ra |\phi \ra = \langle \phi | \langle \psi
_2|
U^\dagger {\bf P}_a^{(1)} U | \psi \vphantom(_1 \ra |\phi \ra . \eqno(5)$$

Note that in the absence of a nonlocal measurement (\ie in the absence
of the operator {\twelveit U}), Eqs.  (4) and (5) would be trivial due
to commutativity of the local operators ${\bf P}_a^{(1)}$ and
$U^{(2)}$.  In general, $U$ does not commute with $U^{(2)}$ or ${\bf
P}_a^{(1)}$.  Thus, Eqs.(4) and (5) represent causality constraints on
the possible nonlocal measurements.  We shall use them below for
deriving necessary properties of nonlocal measurements.

{\bf \chapter {State verification measurements}}

More than half a century after the creation of quantum theory there is
no clear consensus about the interpretation of its basic concept: a
quantum state.  Does it represent some kind of reality or is it just a
mathematical tool for calculating probabilities?  The possibility of
instantaneous verification of a quantum state will manifest its
physical meaning.

We start by investigating the properties of {\it state verification}
measurements.  By a verification of a given state \po ~we understand a
measurement which always yields the answer ``yes" if the measured
system is in the state \po ~and the answer ``no" if the system is in
an orthogonal state \pop.  If the initial state is a superposition of
\po ~and \pop ~then the appropriate probabilities for the answers
``yes" and ``no" will follow from the linearity of quantum theory.

We note that the state verification measurement defined above does not
imply anything about the final state of the system, unlike the
standard quantum measurements in which the final state is an
appropriate eigenstate of the measured operator.  In this sense, state
verification measurement is more basic.

Causality limitations on quantum measurements were used as an argument
against associating physical reality to a quantum
state\rlap.\refmark{\LP}\ Indeed, we will show that causality forbids
performing state verifications using standard quantum measurements
(for example by measuring the projection operator on \po ).
Nevertheless, a method which permits verification of any quantum
(even nonlocal) state was found\rlap.\refmark{\AAV}\ The method is
called ``exchange measurement." \Ref\OP{It is interesting to note that
{\it local} exchange measurements allow to perform local ``perfect
disturbing measurements," see T. Ohira and P. Pearle, {\it Am.  J.
Phys. } {\bf 56} (1988) 692. An explicit procedure for local exchange
measurement was suggested by J.L. Park, {\it Found. Phys.} {\bf 1}
(1970), 23.} The idea of exchange measurement
is to make simultaneous short local interactions with parts of the
measuring device such that the states of the system and the measuring
device will be exchanged.  The novel point in this method is that {\it
local} interactions exchange {\it nonlocal} states.  The result of the
measurement cannot be read by two local observers; the two parts of
the measuring device have to be brought to one place.

Exchange measurements have another very unconventional property: after
the measurement, the system ends up in a state $\vert\psi_{final}\ra$
which is completely independent of the initial state of the system,
but depends only on how the measurement is designed.  Thus, the
exchange measurement has the property of erasing from the system all
information about the initial state of the system.  We emphasize that
this erasing of information takes place at the level of an entire
ensemble of systems.  That is, when exchange measurements are
performed on an ensemble of systems, all the systems in the ensemble
will end up in the same final state $\vert\Psi_{final}\ra$, so that
after the measurement there is no trace of the initial state in the
ensemble.  This is to be contrasted to what happens in the case of
standard quantum measurements.  In the latter, after a measurement,
each individual system in the ensemble ``forgets" its initial state
and ends up in an eigenstate of the measured operator.  The final
ensemble becomes a mixture of eigenstates of the measured operator,
but the probabilities with which these eigenstates mix still reflect
the initial state - they are just the squares of the absolute values
of the projections of the initial state onto the corresponding
eigenstates of the measured operator.

We will show that erasing of information from the system is a generic
property of any causal state verification measurement.  However, not
all information is necessarily erased; the causality requires that
{\it local} information is erased.  The probabilities of the outcomes
of any local measurement performed after the state verification are
independent of the initial state of the system (except possibly for
measurements related to degrees of freedom not involved in the state
verification).

Let us now enunciate the above property in a more precise form.
Consider a measurement designed to verify whether or not a system is
in a given state \po.  By choosing appropriate local orthonormal bases
in part 1 and part 2, we can decompose \po ~as (Schmidt decomposition)
$$
 |\psi_0\ra=\sum_i\alpha_i|i\ra_1|i\ra_2.\eqno(6)
 $$
Let us denote by $H^{(1)}$ and $H^{(2)}$ the Hilbert spaces of part 1
and part 2 respectively, and by $H^{(1)}_0$ and $H^{(2)}_0$ the
subspaces of $H^{(1)}$ and $H^{(2)}$ which are spanned by the basis
vectors $|i\ra_1$ and $|i\ra_2$ corresponding to coefficients $\alpha
_i\neq 0$.  We shall prove that for all initial states \p\ belonging
to the Hilbert space $H^{(1)}_0\!\otimes H^{(2)}_{\vphantom 0}$ the
probability $p(\psi)$ for a result of a local measurement performed on
part 1, after the state verification of \po, has no dependence at all
on the initial state.  In particular, the initial state might be
  \po.  Using notation (1) we
 formulate the following theorem:

\vskip.5cm
 \noindent {\bf
Theorem 1} \par If $|\psi \ra\in H_0^{(1)}\!\otimes H^{(2)}$, then
$p(\psi) = p(\psi _0)$.

\vskip.5cm

In general, however, the initial state is not restricted to
$H^{(1)}_0\!\otimes H^{(2)}_{\vphantom 0}$; it may be any state
belonging to $H^{(1)}\!\otimes H^{(2)}$.  Let us decompose \p ~as
$$
 \P =\alpha \PP+\beta \PPP,\eqno(7)
 $$
 where \pp ~and \ppp ~are the normalized
projections of \p ~onto  $H^{(1)}_0\!\otimes H^{(2)}_{\vphantom 0}$
and onto the
complement $\Big ( H^{(1)}-H^{(1)}_0\Big )\!\otimes H^{(2)}_{\vphantom 0}$
respectively. Then the probabilities of local measurements performed
on part 1
after the state verification measurement may depend on \p ~only via
its
component $\beta \PPP$. We will express this property in the
following theorem:

\vskip.5cm
 \noindent {\bf Theorem 2}

Let $\P=\alpha \PP +\beta \PPP$,\hfill \break
where $\PP\in H_0^{(1)}\!\otimes H^{(2)}$ and $\PPP\in
\Big(H^{(1)}-H_0^{(1)}\Big)\!\otimes H^{(2)}$. Then
$$
 p(\psi)= |\alpha|^2 p(\psi_0)+|\beta|^2p(\psi'').\eqno(8)
 $$

\vskip.5cm

Erasing of local information about the part of the initial state
in $H^{(1)}_0\!\otimes H^{(2)}_{\vphantom 0}$ is the essential
property of the state verification measurement; the sensitivity
of local measurements to \ppp ~and to $\vert\beta\vert$
is trivial.  Indeed, the aim of the state verification measurement is
to distinguish between \po\ and the states orthogonal to it.  But in
order to distinguish between \po\ and the states belonging to
$\big(H^{(1)}-H^{(1)}_0\big)\otimes H^{(2)}_{\vphantom 0}$ it is
enough to perform a local measurement in part 1, and, similarly, a
local measurement performed on part 2 can distinguish between \po and
the states belonging to $H^{(1)}\!\otimes\big(H^{(2)}-H^{(2)}_0\big)$.
Only for distinguishing between \po\ and  other states in the
subspace
$H^{(1)}_0\!\otimes H^{(2)}_0$ is a genuine nonlocal measurement
needed.  But this measurement can be performed by an interaction
applying only to the $H^{(1)}_0\!\otimes H^{(2)}_0$ subspace, not
the complementary subspaces
$\big(H^{(1)}-H^{(1)}_0\big)\!\otimes H^{(2)}$ and
$H^{(1)}\!\otimes\big( H^{(2)}-H^{(2)}_0\big)$.

The rest of this section is devoted to the proof of the above two theorems.

We shall start with the proof of a simple property of state
verification measurement.  As follows from its definition, the
measurement is {\it reliable}, that is, whenever the system is in
$\vert \psi _0\ra$, the answer is always ``yes", while whenever the
system is in an orthogonal state $\vert \psi _{\perp}\ra$, the answer
is always ``no".  Then
$$
 \langle \phi | \langle \psi _{\perp}| U^\dagger {\bf P}_a^{(1)} U | \psi
_0
\ra |\phi \ra = 0 \eqno(9)
 $$
where again $U$ is the unitary transformation describing the state
verification measurement, ${\bf P}_a^{(1)}$ is the projection operator
on a certain outcome of the subsequent local measurement performed in
part 1, and $|\phi \ra$ is the initial state of the measuring device.
Indeed, $ U |\psi _0\ra|\phi\ra$ corresponds to ``yes" states of the
measuring device while $ U|\psi _{\perp }\ra|\phi \ra$ corresponds to
``no" states.  The operator ${\bf P}_a^{(1)}$ does not act on the
states of the measuring device, therefore ${\bf P}_a^{(1)} U |\psi
_0\ra|\phi\ra$ also belongs to a subspace of states corresponding to
the answer ``yes".  Consequently, ${\bf P}_a^{(1)} U |\psi
_0\ra|\phi\ra$ must be orthogonal to $ U|\psi _{\perp }\ra|\phi \ra$.

We proceed now to the proof of Theorem 1 by dividing it into a lemma
and two propositions.
 \vskip .4cm
\noindent {\bf Lemma}
 \par Let $ |\psi \ra = \alpha |\psi_0\ra +
\beta|\psi_{\perp} \ra,~~ \beta \neq 0$, where \pop ~is orthogonal to
\po.

Then, $p(\psi ) = p(\psi_0)$ if and only if $p(\psi_{\perp}) =
p(\psi_0)$.

 \vskip .4cm
\noindent {\it Proof.} Using Eq.(3) we obtain:
$$
 p(\psi) = \langle \phi | \bigl(\alpha^* \langle \psi _0| + \beta^*
\langle
\psi_{\perp}|\bigr)~ U^\dagger {\bf P}_a^{(1)} U ~ \bigl(\alpha |\psi _0\ra
+
\beta |\psi_{\perp} \ra \bigr) |\phi \ra \hfill =~~~~~~~~~~~~~~~~
$$
$$
|\alpha|^2
p(\psi_0) +
|\beta|^2 p(\psi_{\perp}) + \alpha \beta^* \langle \phi | \langle \psi
_{\perp}| U^\dagger {\bf P}_a^{(1)} U | \psi _0\ra |\phi \ra + \alpha^*
\beta
\langle \phi | \langle \psi _0| U^\dagger {\bf P}_a^{(1)} U | \psi _{\perp}
\ra
|\phi \ra . \eqno(10)
$$
Now, from the reliability requirement (9) it follows that the
last two terms vanish, and using the normalization condition $\vert
\alpha\vert^2 +\vert \beta \vert^2 =1$ we, finally, obtain:
 $$
 p(\psi) = p(\psi_0) + |\beta|^2 \bigl(p(\psi _{\perp}) - p(\psi_0)\bigr)
\eqno(11)
 $$
 Thus,
 $p(\psi) = p(\psi_0)$ ~~if and only
if~~ $p(\psi_{\perp}) = p(\psi_0).$

Using the lemma we now prove the following proposition:
\vskip .4cm
\noindent {\bf Proposition 1} \par If the initial state of the system
(prior
to the state verification) can be expressed as a linear superposition

\noindent $$|\psi \ra = \sum_i^N c_iU^{(2)}_i|\psi _0\ra,\eqno(12)$$

\noindent where $U^{(2)}_i$ are unitary transformations in part 2 of the
system, then the probabilities for the results of local measurements performed
in part 1 after the state verification are equal to those obtained if the
initial state were \po :

$$p\big(\sum_i^N c_iU^{(2)}_i\psi _0\bigr) = p(\psi_0) .\eqno(13)$$

\vskip .4cm
\noindent {\it Proof.}~~ We shall prove Eq. (13) by induction on $N$,
the
number of terms in the linear superposition. When $N=1$, Eq. (13) is
true because
it reduces to the causality condition (2). Let us now assume that Eq.(13)
is
true for $N=n$ and let us prove that it holds for $N=n+1$. Let
${[U^{(2)}_{n+1}]}^{-1}$ be the inverse of the unitary operator
$U^{(2)}_{n+1}$. Then, from the causality principle (2), we obtain:
$$
p(\sum_i^{n+1} c_iU^{(2)}_i\psi _0) = p({[U^{(2)}_{n+1}]}^{-1}
\sum_i^{n+1}
c_iU^{(2)}_i\psi _0) = p(\sum_i^n c_i {[U^{(2)}_{n+1}]}^{-1} U^{(2)}_i\psi
_0 +
c_{n+1} \psi_0) . \eqno(14)
$$
Consider now the state ${\cal N}\sum_i^n c_i {[U^{(2)}_{n+1}]}^{-1}
U^{(2)}_i\psi _0$.  Here $\cal N$ is a normalization factor,
appearing because  $\sum_i^n c_i
{[U^{(2)}_{n+1}]}^{-1} U^{(2)}_i\psi _0 + c_{n+1} \psi_0$ is
normalized.  Since the ${[U^{(2)}_{n+1}]}^{-1} U^{(2)}_i$ are unitary
transformations for all {\it i}, it follows from the induction  assumption
 that
$$
 p({\cal N}\sum_i^n c_i {[U^{(2)}_{n+1}]}^{-1} U^{(2)}_i\psi _0) =
p(\psi_0).\eqno(15)
$$
Let us now decompose:
$$
{\cal N}\sum_i^n c_i {[U^{(2)}_{n+1}]}^{-1} U^{(2)}_i |\psi _0 \ra =
\alpha
|\psi_0\ra + \beta|\psi_{\perp}\ra, \eqno(16)
$$
where \pop ~is orthogonal to \po. Then, from Eqs. (15-16)
and the
lemma, it follows that  $p(\psi_{\perp}) = p(\psi_0)$.

Returning now to Eq.  (14), we note that the state appearing in the
last term can be decomposed as
 $$
 \sum_i^n c_i {[U^{(2)}_{n+1}]}^{-1} U^{(2)}_i |\psi _0 \ra
+ c_{n+1} |\psi_0\ra = \Bigl({{\alpha}\over {\cal N}}+c_{n+1}\Bigr) |\psi_0\ra
+ {{\beta}\over {\cal N}} |\psi_{\perp}\ra.
$$
 As we have already established that $p(\psi_{\perp}) =
p(\psi_0)$,
using the lemma again we obtain
$$
 p(\sum_i^n c_i {[U^{(2)}_{n+1}]}^{-1} U^{(2)}_i\psi _0 + c_{n+1} \psi_0)
= p(\psi_0) .\eqno(17)
$$
Inserting (17) in (14) ends the proof of Eq. (13) and of the
proposition.

\bigskip
 To complete the proof of Theorem 1, we have to prove the second
proposition:
\smallskip

\noindent {\bf Proposition 2}

Any state \p\ which belongs to the Hilbert space $H^{(1)}_0\!\otimes
H^{(2)}$ can be expressed in the form of Eq. (12), \ie,
 $|\psi \ra = \sum_i^N c_iU^{(2)}_i|\psi _0\ra$.

\noindent {\it Proof.}
Let $\{|i\ra_1|j\ra_2\}$ be a basis in $H^{(1)}_0\!\otimes H^{(2)}$.
To prove the proposition it is enough to show that by superpositions
of the form of Eq. (12) we can obtain any particular direct product
$|p\ra_1|q\ra_2$.

Consider the unitary transformations $V_1^{(2)}$ and $V_2^{(2)}$ defined by
$$
\eqalign{V_1^{(2)}|p\ra_2=&|q\ra_2\cr
V_1^{(2)}|q\ra_2=&|p\ra_2\cr
V_1^{(2)}|k\ra_2=&|k\ra_2~~~\hbox{\rm for}~~~ k \neq p,q\cr} \eqno(18a)
$$
$$
\eqalign{V_2^{(2)}|p\ra_2=&-|q\ra_2\cr
V_2^{(2)}|q\ra_2=&|p\ra_2~~~\hbox{\rm if}~~~ q \neq p \cr
V_2^{(2)}|k\ra_2=&|k\ra_2~~~\hbox{for}~~~ k \neq p,q \cr}  \eqno(18b)
$$

\noindent Then,
 $$
{1\over {2\alpha_{p}}} V_1^{(2)} |\psi _0\ra - {1\over
{2\alpha_{p}}}V_2^{(2)}|\psi _0\ra=|p\ra_1|q\ra_2,\eqno (19)
$$
\noindent
where $\alpha_{p}$ is the corresponding coefficient in the
Schmidt decomposition (6).  This ends the proof of the proposition.

The proof of the above two propositions completes the proof of the theorem.
Indeed, Proposition 1 says that for any state \p\ which can be
expressed in the
form of Eq. (12) we have  $p(\psi) = p(\psi _0)$, and Proposition 2
says that any state which
belongs to the subspace $H^{(1)}_0\!\otimes H^{(2)}$ can be
expressed in the form of Eq. (12).
 Thus, if $|\psi \ra\in H_0^{(1)}\!\otimes H^{(2)}$,
then
$p(\psi) = p(\psi _0)$.

Using Theorem 1 we will now prove Theorem 2.

\noindent {\it Proof}.~~Using Eq. (3) we obtain
 $$
 \displaylines{ p(\psi) =
\langle \phi | \bigl(\alpha^* \langle \psi '| + \beta^* \langle \psi
''|\bigr)~
U^\dagger {\bf P}_a^{(1)} U~ \bigl(\alpha |\psi '\ra +\beta |\psi '' \ra\bigr)
|\phi \ra =\hfill \cr \hfill |\alpha|^2 p(\psi ') + |\beta|^2 p(\psi '')+
\alpha \beta^* \langle \phi | \langle \psi ''| U^\dagger {\bf P}_a U | \psi
'\ra |\phi \ra + \alpha^* \beta \langle \phi | \langle \psi '| U^\dagger
{\bf
P}_a^{(1)} U | \psi '' \ra |\phi \ra .\hfill (20)\cr}
$$
\noindent
Theorem 1 implies that $p(\psi ') = p(\psi_0)$; therefore,
 we have only to show that the last two terms of Eq. (20)
vanish. Since these terms are complex conjugates of one another, it is
enough
to proof that, say, the first of the two vanishes. Let us
calculate this term using Proposition 2, \ie, the fact that since
$|\psi
'\ra\in H_0^{(1)}\!\otimes H^{(2)}$ it has the form of Eq. (12):
$$
 \alpha \beta^*
\langle \phi | \langle \psi ''| U^\dagger {\bf P}_a U | \psi '\ra |\phi \ra
=
\alpha \beta^* \sum_i^N c_i \langle \phi | \langle \psi ''| U^\dagger {\bf
P}_a
U U^{(2)}_i|\psi _0\ra |\phi \ra . \eqno(21)
$$
Now we shall show that each term in the last sum is equal to zero.
Using the causality principle as in Eq. (5) and taking the
unitary transformation acting on part 2  to be ${[U^{(2)}_i]}^{-1}$,
we obtain:
$$
 \langle \phi | \langle \psi ''| U^\dagger {\bf P}_a U U^{(2)}_i|\psi
_0\ra
|\phi \ra = \langle \phi | \langle \psi ''|U^{(2)}_i
U^\dagger {\bf P}_a^{(1)} U |\psi _0\ra |\phi \ra . \eqno(22)
 $$
Since $|\psi ''\ra\in (H^{(1)}-H_0^{(1)})\!\otimes H^{(2)}$ and
since
${[U^{(2)}_i]}^{-1}$ acts only in part 2 we have also
${[U^{(2)}_i]}^{-1} |\psi ''\ra\in (H^{(1)}-H_0^{(1)})\!\otimes
H^{(2)}$, and therefore the state ${[U^{(2)}_i]}^{-1} |\psi ''\ra $
is
orthogonal to \po. Now,
Eq. (9) implies that the right hand side of Eq. (22) vanishes,
\ie, each
individual term in Eq. (21) vanishes. This ends the proof of Theorem
2.
\vskip 1 cm

{\bf \chapter {Operator measurements}}

We shall now use the results of the previous section in the study of
standard quantum measurements.  The measurement of an operator $A$ can
be considered  a verification measurement of each of its
nondegenerate eigenstates.  It immediately follows that most
operators having some nondegenerate eigenstates are unmeasurable.
Indeed, on one hand, the final state of the system must be locally
independent of its initial state, as follows from Theorems 1 and 2. On
the other hand, if the system is initially in an eigenstate, it should
be undisturbed by the measurement.  Only in very special cases
can these two requirements be simultaneously satisfied.

Let us consider the simplest nonlocal system, two nonidentical spin-1/2
 particles separated in space.  Let \po ~be an arbitrary entangled
state of these particles.  We shall prove that the projection operator
onto \po , ${\bf P}_{\PO}$, is unmeasurable.  Choosing appropriate
local bases, we can write \po\ (Schmidt decomposition)
as
 $$
 \PO = \alpha \u_z \ra \u_{z'} \ra +\beta \d_z \ra
\d_{z'}
\ra ,\eqno(23)
 $$
where $\alpha~,\beta \neq 0$ and where the arrows represent the spin
polarized ``up" or ``down" along some arbitrary directions $z$ and
$z'$.  Consider now two possible initial states ~~~ $\vert \psi_1\ra
=\u_z\ra\d_{z'}\ra$ and $\vert\psi_2\ra=\d_z\ra\u_{z'}\ra$, and let us
suppose that ${\bf P}_{\PO}$ is measurable.  Then, as $\vert\psi_1\ra$
and $\vert\psi_2\ra$ are both eigenstates of $ {\bf P} _{\PO}$
(corresponding to the eigenvalue zero), they must not be disturbed by
the measurement, so the system will end in $\vert \psi_1\ra$ or
$\vert\psi_2\ra$ respectively, which are locally distinguishable.  But
the measurement of $ {\bf P} _{\PO}$ is a verification of \po, and
according to Theorem 1, which applies in this case, it must erase all
local information.  The projection operator $ {\bf P} _{\PO}$ is thus
unmeasurable.

We shall now analyze the measurability of completely
nondegenerate
spin operators. We state our result in the following theorem:

\vskip .4cm \noindent {\bf Theorem 3}

Causality constrains  measurements of nondegenerate spin
operators of a composite system on two spin-1/2 particles such that
the only measurable operators are those with eigenstates  of
two possible types:
$$
\eqalign{|\psi_1\ra =& \u _{z} \ra_{_1} \u _{z'}\ra_{{\vphantom A}_2}\cr
|\psi_2\ra =& \u _{z} \ra_1 \d _{z'}\ra_2\cr |\psi_3\ra =& \d _{z} \ra_1 \u
_{z'}\ra_2\cr |\psi_4\ra =& \d _{z} \ra_1 \d _{z'}\ra_2\cr}\eqno(24a)
$$
or
$$
\eqalign{|\psi_1\ra =&\n \bigl(\u _{z} \ra_1 \u _{z'}\ra_2+ \d _{z} \ra_1
\d
_{z'}\ra_2 \bigr)\cr |\psi_2\ra =&\n \bigl(\u _{z} \ra_1 \u _{z'}\ra_2- \d
_{z}
\ra_1 \d _{z'}\ra_2 \bigr)\cr |\psi_3\ra =&\n \bigl(\u _{z} \ra_1 \d
_{z'}\ra_2+ \d _{z} \ra_1 \u _{z'}\ra_2 \bigr)\cr |\psi_4\ra =&\n \bigl(\u
_{z}
\ra_1 \d _{z'}\ra_2- \d _{z} \ra_1 \u _{z'}\ra_2 \bigr)\cr}\eqno(24b)
$$

The actual eigenvalues are irrelevant;
they must
only be different from each other, so that the operator is completely
nondegenerate.

\vskip .2cm
\noindent {\it Proof.}~~~
Operators of type ($24a$), although they refer to both spins, are
effectively local.  They can be measured by simply measuring the $z$
component of the spin of the first particle and the $z'$ component of
the spin of the second particle.  Operators of  the type
($24b$) are truly nonlocal, as they have entangled eigenstates.
In fact, the eigenstates ($24b$) are all maximally entangled.  The
measurability of these operators has been shown\refmark{\AA,
\AAV} and an explicit measuring method, involving only local
interactions, has been given.  They provided, in fact, the first
example
of nonlocal variables which can be instantaneously measured in the
framework of relativistic quantum mechanics.

What  remains to prove is that if the eigenstates of a
nondegenerate operator cannot be brought to either of the forms
($24a,b$), then causality forbids its measurement.
 Consider first a nondegenerate operator $A$ for which all
its eigenstates are direct products.  Up to an
interchange
of the roles of particle 1 and particle 2, the set of eigenstates
of such an operator can always be written as
$$
\eqalign{|\psi_1\ra =& \u_{z} \ra_1 \u _{z'}\ra_2\cr |\psi_2\ra =& \d_{z}
\ra_1 \u_{z'}\ra_2\cr |\psi_3\ra =& \u_{z''} \ra_1 \d_{z'}\ra_2\cr
|\psi_4\ra
=& \d_{z''} \ra_1 \d _{z'}\ra_2\cr}\eqno(25)
$$
If $z''$ is parallel or antiparallel to $z$ then the set of
eigenstates (25) is equivalent to the set ($24a$).  Let us prove from
causality that indeed, $z''$ must be parallel or antiparallel to $z$.

Let us write
$p(\psi)$
for probability to obtain ${\sigma_z}^{(1)} = -1$  in a
measurement performed on particle 1 immediately
after a measurement of $A$ when $|\psi \ra$ is the
the initial state of the system.  Consider two possible initial states
of the system:
 $$
\eqalign{|\xi_1\ra =&
|\psi_1\ra = \u _{z} \ra_1 \u _{z'}\ra_2,\cr |\xi_2\ra =& \u _{z} \ra_1 \d
_{z'}\ra_2.\cr}\eqno(26)$$
{}From the causality principle it follows that
$$
p(\xi_1) = p(\xi_2) . \eqno(27)
$$
The state $|\xi_1\ra$ is an eigenstate of $A$.  Thus, the measurement
of $A$ does not disturb this state, and, therefore, the probability
to obtain
${\sigma_z}^{(1)} = -1$
 afterwards vanishes,
$p(\xi_1) = 0$.
\eject

On the other hand
 $$
\displaylines{\quad p(\xi_2) = \sum_i |\langle
\xi_2|\psi_i
\ra|^2 p(\psi_i) =\hfill \cr \hfill |\langle \uparrow_ {z} |\uparrow _{z''}
\ra|^2 |\langle \downarrow_ {z} |\uparrow _{z''} \ra|^2 + |\langle
\uparrow_
{z} |\downarrow _{z''} \ra|^2 |\langle \downarrow_ {z} |\downarrow _{z''}
\ra|^2 .\hfill (28)\cr}
 $$
{}From the right hand side of (28) we see that, indeed, $p(\xi_2) = 0$
if, and only if, $z''$ is parallel or antiparallel to $z$.  This ends
the proof that if the eigenstates of a measurable nondegenerate
operator are direct products, they also have
the form ($24a$).

Consider now an operator $A$ which has at least one entangled nondegenerate
eigenstate, say $|\psi_1\ra$. By choosing appropriate local bases we can
write
$$
|\psi_1\ra = \alpha \u _{z} \ra_1 \u _{z'}\ra_2 + \beta \d _{z} \ra_1 \d
_{z'}\ra_2 . \eqno(29)
 $$
We now regard the measurement of $A$ as a verification of the state
$|\psi_1\ra $. For an entangled state both $\alpha $ and $\beta $ are
nonzero,
and Theorem 1 implies that the measurement of A erases from the system all
local information about its initial state. Since the eigenstates of $A$ are
undisturbed by the measurement, they must be locally indistinguishable.
This
requirement can be fulfilled only if $\vert \alpha \vert =\vert \beta
\vert={1\over{\sqrt 2}}$. Indeed, if $\vert \alpha \vert \ne\vert \beta
\vert$
there are no states orthogonal to $\vert\psi_1\ra$ and locally
indistinguishable from it. On the other hand, when $\vert \alpha \vert
=\vert
\beta \vert={1\over{\sqrt 2}}$, the requirement of local
indistinguishability
implies that the eigenstates have the form
 $$
 \vert \psi_i \ra =
{1\over{\sqrt
2}}( \u _{z_i} \ra_1 \u _{z'_i}\ra_2 + e^{i\phi_i} \d _{z_i} \ra_1 \d
_{z'_i}\ra_2) , \eqno(30)
 $$ up to overall irrelevant phases. Note that the
directions $z$ and $z'$ depend on $i$, and they must be chosen such
that the
states $\vert\psi_i\ra$ are mutually orthogonal.

It remains to prove that any four mutually orthogonal states (30) can be
brought, by choosing appropriate local bases, to the form ($24b$).

For simplicity let us first redefine the base vectors such that
$\vert\psi_1
\ra$ reads
 $$
 \vert \psi_1 \ra ={1\over \sqrt{2}} ( \u _z \ra_1
\d_{z'}\ra_2 -
\d _z \ra_1 \u _{z'}\ra_2) . \eqno(31)
 $$
The above form of $\vert\psi_1\ra$ has the property that it is
invariant when we change the basis vectors in an identical way for
both
particles.  Consider now the eigenstate $\vert\psi_2\ra$, which is
orthogonal to $\vert\psi_1\ra$ and locally undistinguishable from it.
Expressed in the same local basis as  (31),
the most general form of such a state (up to an overall phase) is
 $$
\displaylines{\quad \vert\psi_2\ra= {1\over \sqrt{2}}\cos\alpha\bigl(
e^{i{\phi\over 2}} \u_z \ra_1 \u_{z'}\ra_2 + e^{-i{\phi\over2}}\d _z \ra_1
\d
_{z'}\ra_2\bigr) \hfill \cr \hfill~~~~~~ +~ {i\over{\sqrt
2}}\sin\alpha\bigl(
\u_z\ra_1\d_{z'}\ra_2+\d_z\ra_1\u_{z'}\ra_2\bigr). \hfill (32)\cr}
 $$
\noindent
Consider now the local basis transformations given implicitly  by
$$
\eqalign{ \u_z\ra=&e^{-i{\phi\over 2}}\bigl(\cos\beta
\u_{\xi}\ra+i~\sin\beta\d_{\xi}\ra\bigr) \cr \d_z\ra=&e^{i{\phi\over
2}}\bigl(i~\sin\beta \d_{\xi}\ra+\cos \beta\u_{\xi}\ra\bigr)\cr} \eqno(33a)
$$
$$\eqalign{ \u_{z'}\ra=&e^{-i{\phi\over 2}}\bigl(\cos\beta
\u_{\xi'}\ra+i~\sin\beta\d_{\xi'}\ra\bigr)\cr \d_{z'}\ra=&e^{i{\phi\over
2}}\bigl(i~\sin\beta \d_{\xi'}\ra+\cos \beta\u_{\xi'}\ra\bigr),\cr}
\eqno(33b)
$$
 where $\tan2\beta=\cot\alpha$. These transformations preserve
the
form of $\vert\psi_1\ra$,
 $$
 \vert \psi_1 \ra ={1\over \sqrt{2}} ( \u
_{\xi}
\ra_1 \d_{\xi'}\ra_2 - \d _{\xi} \ra_1 \u _{\xi'}\ra_2), \eqno(34)
 $$
and bring $\vert\psi_2\ra$ to the form
 $$
 \vert \psi_2 \ra ={1\over
\sqrt{2}} (
\u _{\xi} \ra_1 \d_{\xi'}\ra_2 + \d _{\xi} \ra_1 \u _{\xi'}\ra_2).
\eqno(35)
 $$
In this new base, the most general form of $\vert\psi_3\ra$
(orthogonal to and locally indistinguishable from $\vert\psi_1\ra$ and
$\vert\psi_2\ra$) is
 $$
\vert \psi_3 \ra = {1\over \sqrt{2}} ( \u _{\xi} \ra_1 \u_{\xi'}\ra_2 +
e^{i\sigma} \d _{\xi} \ra_1 \d _{\xi'}\ra_2).  \eqno(36)
 $$
 By redefining the phases of the base vectors
 $$
\eqalign{ \u_{\xi}\ra
&\rightarrow
e^{i{\sigma\over 2}}\u_{\xi}\ra ,\cr  \d_{\xi}\ra&\rightarrow
e^{-i{\sigma\over
2}}\d_{\xi}\ra \cr}\eqno (37a)
 $$
 $$
\eqalign{ \u_{\xi'}\ra&\rightarrow
e^{i{\sigma\over 2}}\u_{\xi'}\ra , \cr \d_{\xi'}\ra&\rightarrow
e^{-i{\sigma\over 2}}\d_{\xi'}\ra , \cr}\eqno (37b)
 $$
we preserve the form of $\vert\psi_1\ra$ and $\vert\psi_2\ra$ and
eliminate the relative phase between the two terms in (36), and thus
obtain (up to an overall phase)
 $$
 \vert
\psi_3
\ra = {1\over \sqrt{2}} ( \u _{\xi} \ra_1 \u_{\xi'}\ra_2 + \d _{\xi} \ra_1
\d
_{\xi'}\ra_2). \eqno(38)
 $$
 Finally, the eigenstate $\vert \psi_4 \ra$ is
determined  by its orthogonality to $\vert \psi_1 \ra$, $\vert \psi_2 \ra$
and
$\vert \psi_3\ra$, and takes the form $$ \vert \psi_4 \ra = {1\over \sqrt{2}}
(\u _{\xi} \ra_1 \u _{\xi'}\ra_2 - \d _{\xi} \ra_1 \d _{\xi'}\ra_2).
\eqno(39)
$$
This completes our proof, since the set of eigenstates
$\vert\psi_1\ra$,...,$\vert\psi_4\ra$ is, up to
renumbering, equivalent to the set ($24b$).
 \vskip 1.2cm

{\bf \chapter {
Nondemolition verifications and ideal measurements of the
first kind}}
\nobreak

As a final application of Theorem 1, we will study the possibility to
perform {\it ideal measurements of the first kind}.  A basic
assumption in axiomatic nonrelativistic quantum theory\refmark{\Pir}
is that every property of a quantum system may be determined via an
ideal measurement of the first kind.  We will now show that there are
nonlocal properties which cannot be determined in this way.

Ideal measurements of the first kind are a particular case of {\it
nondemolition} measurements.  A {\it nondemolition verification} of
the state \po\ is a state verification measurement with an additional
requirement that, if the result of the measurement is ``yes", then the
system ends up in the state \po.  In particular, if the system is
initially in the state \po, it will remain in this state.  On the
other hand, if the result is ``no", then there are no restrictions on
what will be the final state of the system.  Analogously, we define
{\it verifications of higher-dimensional Hilbert subspaces}: a
verification of a subspace R is a measurement which always yields
``yes" if the state of the system belongs to R and ``no" if the state
is orthogonal to R. As in the case of state verification, no
restrictions are imposed on the state of the system after the
measurement.  A {\it nondemolition} verification of R has the
supplementary property that if the result of the measurement is
``yes", the final state is the projection of the initial state on R.
In particular, if the initial state belongs to R, the state remains
unchanged.

It has already been shown that any state can be verified in a
nondemolition way (exchange measurements\refmark{\AAV}).  Thus, it is
possible to perform ideal measurement of the first kind
to verify an arbitrary state.  However, this is not the case for
verification of higher-dimensional Hilbert spaces.  We will now
present an example of a three-dimensional Hilbert space which cannot
be verified in a nondemolition way, and, therefore, no corresponding
ideal measurement of the first kind is possible.  Consider once again
two nonidentical spin-1/2 particles.  Let R be the subspace of states
which are orthogonal to the state
 $$
 \PO = \alpha \u \ra \u \ra+\beta \d
\ra \d
\ra , ~~~~\alpha,~\beta \neq 0.\eqno(40)
 $$
The proof that R cannot be verified in a nondemolition way is
identical to the proof of the unmeasurability of the projector on \po.
Since \po\ is the unique state orthogonal to R, a nondemolition
verification of R is at the same time a verification (not necessarily
nondemolition) of \po.  Thus, we can apply Theorem 1, \ie, all local
information must be erased.  Consider, however, two possible
initial states
$$
 \vert\psi_1\ra=\u\ra \d\ra ~~,~~~ \vert\psi_2\ra=\d\ra \u\ra.
\eqno(41)
 $$
Both $\vert\psi_1\ra$ and $\vert\psi_2\ra$ belong to R and, therefore,
should be unaffected by the measurement.  But since they are locally
distinguishable, they will lead to locally distinguishable final
states, in contradiction to Theorem 1. This ends our proof.

{\bf \chapter { Conclusions}}

We have proved that even according to the weakest definition of state
verification, requiring only reliability of the measurement, causality
implies that verification of an entangled state must erase local
information.  We have analyzed conditions for which all local
information must be erased by the state verification and have found
that there is a very wide class of such situations (see Theorem 1).
An example is a verification measurement of {\it any} entangled state
of two spin-1/2 particles.  The causality principle states that any
disturbance of a particle just prior to a time $t_0$ cannot affect the
results of local measurements performed on a second particle
immediately after $t_0$.  We, however, have proved the surprising
result that also any disturbance of the {\it second} particle before
$t_0$ does not change probabilities for the results of local
measurements performed on that particle after verification of an
entangled state at $t_0$.

We have also shown in general what
 local information must be erased by
verification of an entangled state (Theorem 2).  These theorems helped
us  analyze the question of measurability of operators.
  We completely analysed the
measurability of nondegenerate spin operators on a system of two
spin-1/2 particle (Theorem 3).  We have shown that causality
imposes
severe constraints: even certain local operators, \ie, operators
with  product eigenstates (but not of the type $(24a)$),
 cannot be measured without violating causality.  However, there
are operators with entangled eigenstates that can be measured
($25b$).  Measurability of all but two types of
operators contradicts the causality principle.  For the operators of
these
two types, there are known measurement procedures that use only
local (and, therefore, causal) interactions.

We applied Theorem 1 to show that for certain Hilbert subspaces, there
is no way to perform an ideal measurement of the first kind without
violating causality.  This raises new difficulties for the
construction of a relativistic axiomatic quantum theory as an
extension of the nonrelativistic one.

We hope that our investigation of the constraints on measuring
nonlocal variables due to relativistic causality can be
extended to more general situations, and that it will lead to
a better understanding of the relativistic quantum theory of
measurement.

After we submitted this manuscript we learned that Bennett et al.
\Ref\BEN{C.H.  Bennett, G. Brassard, C. Crepeau, R. Jozsa, A. Peres,
and W.K.  Wootters, Phys.  Rev.  Lett., {\bf70}, 1895 (1993).} have
found a method for teleportation of quantum states.  A similar method
can serve as an alternative to nonlocal exchange measurements.  In
this method the local information is also completely erased in
accordance with our Theorem 1.

 \vskip 1cm

\centerline{\bf Acknowledgements}

It is a pleasure to thank Yakir Aharonov and Daniel Rohrlich for very
helpful discussions.  We are thankful to the Referee for pointing out
relevant references.  The research was supported in part by grant
425/91-1 of the the Basic Research Foundation (administered by the
Israel Academy of Sciences and Humanities).

\endpage
\refout

\bye